\documentclass{pasj00}
\begin{document}
\SetRunningHead{J. G. Coelho, and M. Malheiro}{Magnetic Dipole Moment of SGRs and AXPs Described as Massive and Magnetic White Dwarfs}

\title{Magnetic Dipole Moment of SGRs and AXPs\\ Described as Massive and Magnetic White Dwarfs}

\author{Jaziel G. \textsc{Coelho} and Manuel \textsc{Malheiro} %
}
\affil{Departamento de F\'{i}sica, Instituto Tecnol\'{o}gico de Aeron\'{a}utica, CTA,
S\~{a}o Jos\'{e} dos Campos, 12.228-900, Brasil}
\email{jaziel@ita.br}


\maketitle

\begin{abstract}
The Anomalous X-ray Pulsars (AXPs) and Soft Gamma-ray Repeaters (SGRs) are
some of the most interesting groups of pulsars that have been intensively studied
in the recent years. They are understood as neutron stars (NSs) with super strong magnetic fields, namely $B\gtrsim10^{14}$ G. However, in the last two
years two SGRs with low magnetic fields $B\sim(10^{12}-10^{13})$ G have been detected. Moreover, three fast and
very {\it magnetic} white dwarfs (WDs) have also been observed in the last years. Based on these new pulsar
discoveries, we compare and contrast the magnetic fields, magnetic dipole moment, characteristic ages, and X-ray steady luminosities of these two
SGRs (in the WD model) with three fast white dwarfs, to conclude that they show strong similarities corroborating an alternative description of several
SGRs/AXPs as very massive and magnetic white dwarfs.
The pulsar magnetic dipole moment $m$ depending only on the momentum of inertia $I$,
and observational properties, such as the period $P$ and its first time derivative $\dot{P}$, can help to identify the scale of
$I$ for SGRs/AXPs. We analyze the pulsar magnetic dipole moment $m$ of SGRs and AXPs when a model based on a
massive fast rotating highly magnetized white dwarf is considered. We show that the values for $m$ obtained for several SGRs 
and AXPs are in agreement with the observed range $10^{34}{\rm emu}\leq m \leq10^{36}{\rm emu}$ of 
isolated and polar magnetic white dwarfs. This result together with the fact
that for {\it magnetic} white dwarfs $B\sim(10^6-10^8)$ G their magnetic dipole moments are almost independent of 
the star rotation period ($10^{4}\lesssim P \lesssim10^{6} {\rm s}$) - a phenomenology not shared by neutron stars
pulsars - suggests a possible {\it magnetic} white dwarf nature for some of SGRs/AXPs that have much smaller periods ($P\sim 10$ s).
Moreover, since for pulsars the dipole radiation power is proportional only to $m$ and to the rotational star frequency, we can explain
in the WD model - considering only the different scales of the magnetic dipole moment for WDs and NSs - why the steady luminosity $L_X$ for several 
SGRs/AXPs (and in particular the low-$B$ SGRs) compared
to those of X-ray Dim isolated neutron stars (XDINs) and high-$B$ pulsars obey the ratio
${L_X}^{\rm SGRs/AXPs}/{L_X}^{\rm XDINs}\sim m_{\rm WD}/m_{\rm NS}\sim10^3$: all these X-ray sources have essentially the 
same rotational periods ($P\sim10$ s) and the X-ray luminosity is correlated to the spin-down luminosity which is equal to the 
dipole radiation power in the dipole model.
\end{abstract}
\newpage

\section{Introduction}
Over the last decade, observational evidence has suggested that Soft Gamma-ray Repeaters (SGRs) and Anomalous X-ray Pulsars (AXPs)
belong to a particular class of pulsars. The SGRs/AXPs are understood as slowly rotating neutron stars (NSs) that, 
in contrast to rotation powered radio pulsars and accretion powered X-ray pulsars, are not powered by their spin-down energy losses,
but by the energy stored in their extremely large magnetic fields $B\gtrsim10^{14}$ G.
They are known as very slow rotating isolated pulsars with rotational periods
in the range of $P\sim(2-12)$ s, a narrow range comparing to ordinary pulsars; and spin-down rates
of $\dot{P}\sim(10^{-13}-10^{-10})$ s/s, in contrast to $\dot{P}\sim(10^{-15}-10^{-14})$ s/s for ordinary pulsars.
Their persistent X-ray luminosity, as well as the bursts and flares typical
of these sources, are believed to be powered by the decay of
their ultra strong magnetic fields (see Mereghetti 2008 for review).

SGRs are observed with their bright and short bursts of soft $\gamma$-rays and X-ray radiation and
hence they are considered as a subclass of gamma-ray bursts (GRBs). Moreover, their generally large spin-down rates, 
strong outburst energies of $\sim(10^{41}-10^{43})$ erg and giant flares of $\sim(10^{44}-10^{47})$ erg, make them 
different from the ordinary pulsars. However, the giant flares have been observed so far only 
from SGRs. AXPs, on the other hand, are distinguished from X-ray binaries by their narrow period distribution, 
soft X-ray spectrum, faint optical counterparts, and long term spin-down.
However, observations performed over the last few years have led to new discoveries pointing out many
similarities between these two classes of sources. The magnetar model, firstly developed for the SGRs, has also been applied to
the AXPs and hence they are often classified together, suggesting that AXPs and SGRs belong to the same family.
\newpage
{\bf\emph{The Magnetar Model}.}
In the twisted magnetosphere model of magnetars, the observed X-ray luminosity $L_X$ is determined both by its
surface temperature and by magnetospheric currents, the latter due to the twisted dipolar
field structure (see~\cite{Duncan}; \cite{Thompson}).
The luminosity from the hard tails observed in the X-ray spectra contributes significantly to the total energy output 
in AXPs/SGRs, indicating the presence of non-thermal phenomena in the magnetosphere of underlying AXPs/SGRs.
The surface temperature in turn is determined by the energy output from within the star due to magnetic field decay, 
as well as by the nature of the atmosphere and the stellar magnetic field strength. This surface
thermal emission is resonantly scattered by the magnetospheric current, thus resulting in an overall spectrum similar to a Comptonized
blackbody (see e.g.,~\cite{Beloborodov};~\cite{NandaRea2008}). In addition, the surface heating by return currents
is believed to contribute substantially to $L_X$, at least at the same level as the thermal component induced from the 
interior field decay. Magnetar outbursts in this picture occur with sudden increases in twist angle, consistent with
the generic hardening of magnetar spectra during outbursts (e.g.,~\cite{Kaspi};~\cite{Woods}). 

Furthermore, the strong magnetic field explains the confinement of the hot plasma, required for the subsequent tail with a softer spectrum
pulsating at the NS rotation period, and the short bursts in almost all AXPs/SGRs with peak luminosity exceeding 
the Eddington limit for a NS by a few orders of magnitude, and high frequency quasi-periodic oscillations (QPOs) (see~\cite{Mereghetti}).
The origin of the quasi-periodic oscillations (QPOs) observed in the giant flares of soft gamma-ray repeaters (SGRs) remains uncertain.
Current models explore the idea that long-term quasi-periodic oscillations are trapped at the turning points of the continuum 
of torsional magneto-elastic oscillations in the magnetar's interior. Recently, efforts have been made
using two-dimensional, general-relativistic, magneto-hydrodynamical simulations, coupled to the evolution of shear waves
in the solid crust, in order to explore the viability of this model when a purely dipolar magnetic field is
assumed (see~\cite{Gabler2012};~\cite{Gabler2013}). They showed that axisymmetric, torsional, magneto-elastic oscillations of magnetars with
a superfluid core can explain the whole range of observed quasi-periodic oscillations in the giant flares
of soft gamma-ray repeaters. There exist constant phase, magneto-elastic QPOs at both low ($f<150$Hz) and
high frequencies ($f>500$Hz), in full agreement with observations. The range of magnetic field strengths
required to match the observed QPO frequencies agrees with that from spin-down estimates. 

The recent discovery of radio-pulsed emission in four of this class of sources, where the spin-down rotational energy lost
$\dot{E}_{\rm rot}$ is larger than the X-ray luminosity $L_X$ during the quiescent state - as in normal pulsars - opens the 
question of the nature of these radio sources in comparison to the other SGRs/AXPs (we have also recently studied 
this point in~\cite{jaziel}).
According to the fundamental plane for magnetars, a plot of $L_X$ versus the spin-down luminosity (see~\cite{Turolla} for details),
four of a total of about 20 SGRs/AXPs should have radio-pulsed emission:
XTE J1810-197, 1E 1547.0-5408, PSR J1622-4950, and SGR 1627-41. Basically, is propose that the magnetar
radio activity can be predicted from the knowledge of the star's
rotational period, its time derivative, and the quiescent X-ray luminosity, when $L_X/\dot{E}<1$. It is true that for one of these sources,
SGR 1627-47, no radio emission has been
detected yet, because it is unfavorably affected by distance, scattering, or lack of sensitive
observations at the time its pulsed radio emission was possibly
expected to be brighter. Furthermore, the discovery of 3.76 s pulsations from the new burst source at the Galactic center,
using data obtained with the {\it NuSTAR} Observatory was recently reported. SGR J1745−29 is the fourth magnetar detected in radio wavelengths, 
very similar to the other radio SGRs/AXPs, where also $L_X/\dot{E}\simeq0.02$ is less than one.
The {\it Swift} satellite has also observed the sudden turn-on of a new radio source near Sgr A*. This result, combined with the
detection of a short hard X-ray burst from a position consistent with the new radio SGR, suggests that this source is in fact
a new SGR in the Galactic Center. The combination of a magnetar-like burst, periodicity and
spectrum led to the identification of the transient as a likely new magnetar in outburst (see the recent works
about this new SGR in~\cite{Kennea};~\cite{Kaya}).
However, the pulsar's unusually large Faraday rotation used to estimate the magnetic
field indicates a quite low value $B\gg50$ $\rm\mu G$, in comparison with all the others magnetars. 
Extra information about the gas in the central 10 pc of the Galactic Center must be used for a more robust estimate
of the magnetic field (see~\cite{Nature2013}).
The condition $L_X/\dot{E}<1$ proposed in Rea et al. (2012) that magnetars need to satisfy in order to
explain the radio emission for some of these sources, in our understanding
it is not well justified in the magnetar model. The basic assumption
of this model is that steady X-ray luminosity and not only X-ray burst and flares are powered by ultra strong magnetic fields of the star. Thus, 
the connection proposed between the X-ray steady luminosity and the spin-down rate of the rotational energy 
for the radio magnetars seems to be in contradiction with the basic assumptions of this model, where these two quantities are not correlated.

It is appropriate to recall now a few other difficulties of the magnetar model in fitting observations, following 
Malheiro, Rueda and Ruffini (2012) (see references therein). In particular,
e.g.: (1) as recalled by Mereghetti (2008), \textquotedblleft up to now, attempts
to estimate the magnetic field strength through the measurement
of cyclotron resonance features, as successfully done for accreting pulsars, 
have been inconclusive\textquotedblright; (2) the prediction of the
high-energy gamma ray emission expected in the magnetars has
been found to be inconsistent with the recent observation of the
Fermi satellite (see e.g.,~\cite{Tong2010};~\cite{Tong2011}); (3) it has been shown that the attempt to relate magnetars to 
the energies of the supernova remnants (see e.g.,~\cite{Allen};~\cite{Ferrario2006};~\cite{Vink2006};~\cite{Vink2008}) 
or the formation of black holes is not viable (see~\cite{kasen}).

{\bf\emph{Alternative Models}.}
Thus, even if the magnetar model has been quite successful in explaining the phenomenology of SGRs/AXPs, no conclusive, 
direct measure of the surface magnetic field has been claimed yet. Furthermore, we cannot ignore
that the recent astronomical observations of old SGRs (characteristic age $\sim(10^6-10^7)$ Myr) with low surface magnetic field (see Section 3),
as well as the four SGRs/AXPs showing radio emission pointed out above,
have opened space for alternative models, such as the possible presence of a fallback disk slowing down the neutron star pulsar up to the current
spin period (see e. g.,~\cite{Alpar2011};~\cite{Trumper}), or more exotic scenarios involving the hypothesis of quark
stars (see e.g.,~\cite{Itoh1970};~\cite{Olinto};~\cite{Dey1998};~\cite{MMalheiro2003};~\cite{Horvath2007};~\cite{Paulucci};~\cite{Negreiros2009}) 
to explain these types of sources (see e.g.,~\cite{Xu2007}).

{\bf\emph{The White Dwarf model}.}
An alternative description of the SGRs/AXPs
based on rotating highly magnetized and very massive white dwarfs has been proposed recently by Malheiro, Rueda and Ruffini (2012), 
following previous work by Morini et al. (1988) and Paczynski (1990). 
Moreover, observations of massive fast rotating highly magnetized white dwarfs by dedicated missions as that of the 
X-ray Japanese satellite Suzaku (see e.g.,~\cite{Terada2};~\cite{Terada3};~\cite{Terada4};~\cite{Terada5}) 
has led to a confirmation of the existence of WDs sharing common properties with NS pulsars,
hence called WD pulsars (see~\cite{Terada2013}, for a latest review about white dwarf pulsars).
In this new description several observational features are easy understood and well explained as a consequence of the large radius of a massive
white dwarf, which manifests itself as a new scale of mass density, moment of inertia, rotational energy, and magnetic dipole momentum in
comparison with the case of neutron stars:

a) The existence of stable WDs can explain, the range of long rotation periods
$2\lesssim P\lesssim 12$ seconds observed in SGRs and AXPs. In particular,
the fact we do not observe any SGRs/AXPs with $P\leq 2$ s is 
a consequence of the low surface gravity and density of a white dwarf compared with
those of neutron stars. Boshkayev et al. (2012) performed
a self consistent calculation of dense and very fast rotating white dwarfs and
obtained the minimum rotational periods $\sim 0.3$, $0.5$, $0.7$ and $2.2$ seconds
for $^4\rm He, ^{12}\rm C,^{16}\rm O$, and $^{56}\rm Fe$ WDs respectively. 

b) The long standing puzzle of the energetic balance of the SGRs and AXPs pulsars is solved:
the steady X-ray luminosity $L_X$ observed
is smaller than the loss of rotational energy of the white dwarf, $L_X <\dot{E}_{\rm rot}$,
because $\dot{E}_{\rm rot}$ is
$\sim10^5$ orders of magnitude larger due to the WD moment of inertia in
comparison with the neutron star one. SGRs and AXPs are rotation-powered massive white dwarf pulsars
in complete analogy with neutron star pulsars.

c) The large steady luminosity $L_X\sim10^{35}$ erg/s, for such slow pulsars
($\Omega\sim 1$ Hz) is also understood,
as a consequence of the large radius of the dense WD that produces a large magnetic dipole moment,
in the range of these magnetic WDs, as we will discuss in this paper.
It has usually been thought that white dwarfs could also behave as pulsar since their magnetic fields could be sufficiently strong to 
produce pulsar emission. In the white dwarf model, since SGRs/AXPs are magnetic white dwarfs, they are
spinning too fast, close to the Kepler frequency, and not slow as in the case of neutron stars.
These high WD rotational frequencies together with their large radii in comparison to the NS ones, will 
produce strong potential energy differences, if they are very magnetized WD, and will be able to emitted from X-rays to
even Gamma-rays (see~\cite{Kashiyama}).
As we will see in the Section 3, several WDs were recently observed with magnetic fields up to $\sim10^9$ G, and if SGRs/AXPs are 
white dwarfs, the $B$ field are even larger $\sim(10^9-10^{10})$ G, large enough to power the pulsar emission.

d) The large $\dot{P}$ observed ($10^{-10}-10^{-13}$ s/s) are also a consequence of the large WD
magnetic dipole moment, as well as of the large radius and the momentum of inertia, and not only
of the magnetic field, as is the case for neutron stars
in the magnetar model. Overcritical magnetic fields of the order of $(10^{14}-10^{15})$ G are no longer
needed to explain the large spin-down breaking of the SGRs and AXPs pulsars: they can be understood
as highly magnetized WDs with large magnetic fields ($10^8-10^{10}$) G. Magnetic white dwarfs
tend to be more massive and smaller (\cite{Tout}), exactly
the properties shared by SGRs and AXPs as rotationally powered dense
white dwarfs (\cite{MMalheiro}). Recent studies predicted that the masses of high-field magnetic 
WDs should be larger than the average mass of WDs (see e.g.,~\cite{Silvestri};~\cite{Kawka}). 

e) The burst activity quite frequent in these sources is a consequence of their
large angular velocity, near to break-up and close to rotational instability.
In this critical situation, gravity can stress the star, producing glitches (associated sudden shortening of the period) and giving
origin to the outburst and large flares observed (\cite{Kaspi}; \cite{Woods}). The scale of the rotational energy
delivery by these glitches is also in agreement with the outburst and giant flare
energies observed of $(10^{41}-10^{46})$ erg, since the WD rotational energies
is at least $10^4$ larger than that of the NS due to the large momentum of inertia (see Figs. (4) and (7)
of Malheiro, Rueda and Ruffini 2012).
The occurrence of these glitches 
can be explained by the release of gravitational energy associated with a sudden contraction and
decrease of the moment of inertia of the white dwarf,
consistent with the conservation of angular momentum. The energetics of steady emission as well as that
of the outbursts following the glitch can be simply
explained in term of the loss of rotational energy, in
view of the moment of inertia of the white dwarfs, being much
larger than that of neutron stars.

f) The small population of SGRs and AXPs observed, only $23$ in more than 1800 pulsars, is also understood, since
these WD pulsars are quite fast and very magnetic, and as a consequence rarely formed (\cite{Ilkov}).
Moreover, astronomical observations indicate that isolated magnetic white dwarfs
with high magnetic field are only 10$\%$ of the total magnetic stars found (see \cite{Tout}). Recently, the Sloan Digital Sky Survey (SDSS)
found six very massive white dwarfs with $B\gtrsim10^8$ G (see \cite{Kepler}).

Furthermore, as explained in Malheiro, Rueda and Ruffini (2012), a possible formation mechanism of fast WDs originates in
a binary system composed of a WD and a late evolved companion star, close to the process of gravitational collapse.
We cannot discard the possibility that a WD can be captured by a star cluster to
form such a binary system. Moreover, according to the McGill SGR/AXP Online Catalog, there
are currently twenty confirmed magnetars, consisting of nine
soft gamma repeaters and eleven Anomalous X-ray
Pulsars, as well as six magnetar candidates (four
SGR and two AXP candidates). These numbers are small
relative to the total number of known rotation-powered pulsars,
nearly 2000, according to the online ATNF Pulsar Catalog (see~\cite{Manchester}).
Thus, this uncommon phenomena is consistent with the possibility of a rare capture.

The origin of fast white dwarfs generated by supernovae Ia has also been investigated, where the
rapidly rotating WD is formed shortly after the stellar formation episode, and the delay from stellar formation to explosion
is basically determined by the spin-down time of the rapidly rotating merger remnant (see e.g., \cite{Ilkov}).
It has been only shown recently that high-field magnetic white dwarfs (HFMWDs) might be the result of white dwarf
mergers as was long-suspected (see~\cite{Wickramasinghe}). Rueda et al. (2013) showed that the merger of a double degenerate system can explain the
characteristics of the magnetar AXP 4U 0142+61, consistent with an approximate 1.2 $M_\odot$ white dwarf,
resulting from the merger of two otherwise ordinary white dwarfs of masses 0.6 $M_\odot$ and 1.0 $M_\odot$, 
surrounded by the heavy accretion disk produced during the merger.

It is worth pointing out, before opening a discussion concerning difficulties of the white dwarf model in fitting observations,
that the description 
of SGRs/AXPs as very massive and magnetized white dwarfs ($B\sim 10^8-10^{10}$ G) is quite plausible and can be seen
as a natural extrapolation of the star surface magnetic field of magnetized and isolated white dwarfs recently discovered,
with $B$ from $10^7$ G all the way up to $10^9$ G (see the recent work of~\cite{Kepler}).
It is also worth mentioning the fact that most of the observed magnetized
white dwarfs are massive; see e.g. REJ 0317-853 with $M\sim$ $1.35 M_\odot$ and $B\sim(1.7-6.6)\times10^8$ G 
(see e.g.,~\cite{Barstow};~\cite{Kulebi}); PG 1658+441 with $M\sim$ $1.31M_\odot$
and $B\sim2.3\times10^6$ G (see e.g.,~\cite{Liebert1983};~\cite{Schmidt1992}), 
and PG 1031+234 with the highest magnetic field $\sim10^9$ G (see e.g.,~\cite{Schmidt1986};~\cite{Kulebi2009}).

It is also appropriate to recall the main difficulties concerning the white dwarf model, in order to explain some of the
phenomenology of SGRs/AXPs:

a) {\it Association of Supernova Remnants (SNR)} in two or three of the magnetars, is usually advocated as an indication of a neutron star nature
for these sources, since SNR are considered clear signs of the collapse of a massive star. The formation mechanism of fast WDs based on
a binary system proposed by Malheiro, Rueda and Ruffini (2012), is compatible with a SNR due to the collapse of the late evolved WD companion star.
Furthermore, the magnetar association to SNR is only robust in three cases (see~\cite{Gaensler}), but not considered significant in several other
cases that were proposed in the past (see e.g.,~\cite{Mereghetti}).
The case of the association of AXP 1E 2259+586 with SNR G109.1-1.0 (CTB 109) was analyzed by Paczynski (1990) assuming the merger of a binary
system of an ordinary white dwarf of mass $\sim(0.7-1)M_\odot$, leading both to the formation of a fast rotating white dwarf, and to the supernova
remnant (see section 8 of~\cite{MMalheiro}).
The second case concerns the AXP 1E 1841-045 associated with the SNR Kes 73 (see~\cite{Vasisht}) where 
a pulsar PSR J1841-0500 was recently found located
at only 4' from this AXP, supporting the binary scenario discussed before (see Note added after submission of~\cite{MMalheiro}). Thus, the
existence  of SNR near to the SGRs/AXPs cannot be used to clearly exclude a white dwarf nature of these sources.
The third case concerns the AXP 1E 1547.0-5408 sited at the center
of SNR G327.24-0.13 (see e.g.~\cite{Gelfand}), which is one of the magnetars that emitted in radio as discussed before.
In our understanding, this source is in fact a neutron star pulsar, as are the others
radio magnetars presented recently in Coelho \& Malheiro (2013), and cannot be identified with a massive white dwarf. Its outburst activity, well
studied by Bernardine et al. (2011), is due to the high surface magnetic field of this star, similar to the magnetospheric activity also observed
in the Sun. We claim that this source is a high-$B$ pulsar, but not a magnetar, since its steady luminosity can be well explained in terms of 
the neutron star spin-down rate energy ($L_X<\dot{E}_{\rm rot}$), and does not require magnetic energy to produce this luminosity.

b) {\it SGRs and AXPs are associated with massive, young star clusters, but white dwarfs are usually old.} Recently, two SGRs with low magnetic
fields were found with characteristic ages much higher (Myr) than those of the others (see McGill catalog). The few possible associations
of SGRs/AXPs with clusters of massive stars does not exclude the possibility that a white dwarf
can be captured by a star cluster and
form the binary system that will give origin to the fast white dwarf formation mechanism already discussed. Since, several AXPs are not located
in young star clusters, the merger of two usually older white dwarfs, can also explain the small characteristic ages of these sources. As we already
discussed, recently~\cite{Rueda2013} showed that the merger of a double degenerate system can explain the characteristic age of the AXP 4U 0142+61,
$\sim60$ kyr, consistent with the characteristic ages of several magnetars.

c) {\it The magnetic fields of white dwarfs are not strong enough to produce pulsar emission at the slow periods associated with AXPs.} We have already 
commented on this point before and will dedicate more discussions to it when we present the values for the dipole magnetic moments of
isolated and very magnetic white dwarfs. Kashiyama et al. (2011) address this point in their paper, and showed that because of the large radii
and momenta of inertia of WDs, magnetic fields of $B\sim10^6-10^{10}$ G (exactly the ones obtained for SGRs/AXPs), can produce
high potential energies as in normal neutron star pulsars in order to explain their large emission, from X-rays up to gamma-rays.

d) {\it AXPs have been shown to vary their X-ray luminosity. In particular the transient AXPs seem to contradict the assumption
that quiescent X-ray luminosity is ascribed to spin-down luminosity}. As explained in Mereghetti (2008), in the best theory available, the long-term
variations of X-ray emission on magnetars comes from currents supported in twisted magnetosphere or sudden reconfigurations of the
magnetosphere when unstable conditions are reached. The star heated in these events will pass by through a subsequent cooling process
that could explain the observed long-term decays in the soft X-ray emissions
of AXPs. Thus, the origin of the X-ray variability comes essentially from magnetospheric effects and reconnections expected to occur on very
magnetized WDs. This is not correlated directly to the quiescent X-ray luminosity that is associated
to the spin-down luminosity for the case of the white dwarf model. Furthermore, it is interesting to pointed out that almost all the transient
AXPs that show a large luminosity range variability are the some that emit in radio (except the AXP CXO J164710.2-455216 in the massive star cluster
Westerlund), exactly the ones that we have already discussed, for which $L_X<\dot{E}_{\rm rot}$, with $\dot{E}_{\rm rot}$ calculated as neutron stars. 
This is a strong indication that the energy source for the
X-ray variability and the steady luminosity have different origins: the first due to the large magnetic fields of 
the magnetosphere (this is expected to also happen in the magnetosphere of very magnetized rotating white dwarfs, as
found in the transient radio source GCRT J1745-3009 by~\cite{Zhang}), and the second from the spin-down luminosity
(only possible in the white dwarf model, with the exception of the radio AXPs that, as discussed in~\cite{jaziel},
obey a linear log-log relation between $L_X$ and $\dot{E}_{\rm rot}$, as expected for neutron star pulsars).

e) {\it The quasi-periodic oscillations (QPOs) at frequencies almost up to 1 kHz in SGRs following giant flares, are
very high-frequency oscillations to be supported by a white dwarf}. As we already discussed, axisymmetric, torsional, 
magneto-elastic oscillations of neutron stars can 
explain these QPOs. However, we do not observe directly the vibration
of the crust, but only their effect on the X-ray emission. This is testified by the sporadic nature of the observed signals, that could be originated
by the geometry of the magnetic field and its complex effect on the radiation beam patterns (see~\cite{Mereghetti}). Thus, the 
origin of the QPOs is possibly not associated with magneto-elastic oscillations of the star surface, but instead
with more complex process due to the hot plasma of the star and magnetospheric oscillations,
since the giant flare tail emission is thought to originate in
the fraction of the energy released in the initial spike that remains trapped in the star magnetosphere, forming an optically
thick photon-pair plasma. The same effect for very magnetized white dwarfs, optically thick pairs in equilibrium with radiation trapped by magnetic fields
may also be present in the magnetosphere if the WD magnetic fields are high $B\sim10^9-10^{11}$ G, in the range of the values obtained
for SGRs/AXPs in the white
dwarf model. Furthermore, as we already pointed out, pair production activities in the magnetosphere of a rotating white dwarf
have already been observed in the transient radio source GCRT J1745-3009 by Zhang \& Gil (2005), and also in 
the AE Aquarii (see e.g.,~\cite{Terada2};~\cite{Terada3};~\cite{Terada4};~\cite{Kashiyama}). Thus, we cannot exclude the 
high frequencies of the QPOs from occurring in very massive and fast white dwarfs if they are very magnetized.

f) {\it Hard X-ray emission (up to 1 MeV) which is highly pulsed and at a luminosity that is often more than the soft X-ray luminosity}.
Such a spectral feature is clearly already in evidence for rotating white dwarfs, in particular the WD AE Aquarii observed by the Suzaku satellite,
following the work of Terada et al. 2008c. It was concluded, as presented in more detail in Malheiro, Rueda and Ruffini (2012), that the hard X-ray pulsations
observed on this white dwarf, in addition to the thermal modulation of the softer X-ray band, should have no thermal origin, but possibly the Synchrotron
emission with sub MeV electrons. Since the magnetic fields for almost all SGRs/AXPs as white dwarfs are at least two orders of magnitude higher
than the one of AE Aquarii (see Table~1) we would expected an even harder X-ray emission spectrum up to MeV.

In this work, we will discuss the magnetic dipole moment $m$ of neutron stars and magnetic white
dwarfs, to stress that the values obtained for $m$ for SGRs and AXPs as white dwarfs are in 
agreement with those of polar and isolated magnetic WDs. Using the catalog of the Sloan Digital Sky Survey (SDSS) project, 
we selected from the sample of 480 white dwarfs that have high magnetic field strength in the range of $10^4$ to
$10^9$ G (\cite{Kawka}), 82 WDs for which both period $P$ and magnetic field $B$ are known. 
These recent astronomical observations of white dwarfs allow us to conclude that the
range of the magnetic dipole moment of polar and isolated {\it magnetic} white dwarfs is
$10^{34}{\rm emu}\leq m \leq10^{36}{\rm emu}$ and also almost independent of the star period. This systematic study of magnetic
white dwarfs applied to SGRs/AXPs described as white dwarf pulsars was not performed in our previous works (\cite{MMalheiro};~\cite{Coelho}).
For the first time, we will show that SGRs/AXPs as white dwarf pulsars also have the same magnetic dipole moment of the magnetic WDs
even with much small periods ($P\sim10$ s). Thus, the
astronomical observation that the WD magnetic dipole moment for polar and isolated white
dwarfs is always in the range indicated before, despite the fact that their rotational periods change from $10^4$ to $10^6$ s, is an important evidence
that also suggests the WD nature of SGRs/AXPs, since they share these same properties. 
For both millisecond and long period NSs pulsars,
we do not see this phenomenology, since they have quite different magnetic moments that increase with the period.
The larger values of the magnetic dipole moment for a white dwarf compared to those of neutron star X-ray pulsars 
($10^{28}\leq m\leq 10^{31}$ emu), can explain for instance the large steady luminosity $L_ X$ seen in SGRs/AXPs of $L_X\sim10^{35}$ erg/s, for very
slow pulsars rotational frequencies ($\Omega\sim 1$ Hz).
Furthermore, we extend the comparison done in Malheiro, Rueda and Ruffini (2012) of the observational 
properties of one low-$B$ SGR and one fast white dwarf,
including the recent observations of one more low-$B$ SGR and two other fast, magnetic white dwarfs.
We conclude that several features of these two SGRs with low magnetic fields 
are still very similar to those of fast, magnetic white dwarfs, an important result that shows some of the SGRs/AXPs, 
at least these SGRs with low-$B$, could be identified with very fast and magnetic massive white dwarfs.

{\bf This work is organized as follow:} in section 2, we discuss the standard magnetic dipole model for rotation-powered pulsars in order
to show the scale of the dipole magnetic moment in WDs and NSs. In section 3, we present the
recent discoveries of SGR 0418+5729 (see \cite{NandaRea}) and Swift J1822.3-1606 (see \cite{NandaRea2}; \cite{Livingstone})
with low magnetic field sharing some properties (\cite{Coelho}; \cite{Coelho2}) with the recent detected fast WD pulsars 
AE Aquarii and RXJ 0648.0-4418, and the candidate EUVE J0317-855, supporting the description of some SGRs/AXPs as
white dwarf pulsars. In section 3, we present the WDs magnetic dipole moments in comparison to the NSs ones, suggesting
the possibility of some SGRs/AXPs belonging to a class of very fast, magnetic massive white dwarfs. Finally, in section 4 we 
summarize our conclusions.

\section{The standard magnetic dipole model for rotation-powered pulsars}

The magnetic dipole moment is related to the magnetic field strength at the magnetic pole of the star $B_p$ by,
\begin{equation}\label{magnetic_moment}
\mid\overrightarrow{m}\mid= \frac{B_p R^3}{2},
\end{equation}
where $R$ is the star radius. If the star magnetic dipole moment is misaligned with the spin axis
by an angle $\alpha$, the energy per second emitted by the rotating magnetic dipole is 
(see e.g., \cite{Shapiro} and references therein),
\begin{equation}\label{edot}
\dot{E}_{\rm dip}= -\frac{2}{3c^3}\mid \ddot{m}\mid^2= -\frac{2\mid\overrightarrow{m}\mid^2}{3c^3}\Omega^4\rm sin^2\alpha,
\end{equation}
where $\ddot{m}$ is the second derivative of the magnetic dipole moment, $\Omega=2\pi/P$ its rotational frequency and $c$ the speed of light.
Thus, the physical quantity that
dictates the scale of the electromagnetic radiated power emitted is, together
with the angular rotational frequency, the magnetic dipole moment $m$ of the star. The fundamental physical idea of the rotation-powered
pulsar is that the X-ray luminosity - produced by the dipole field - can be expressed
as originated from the loss of rotational energy of the pulsar,
\begin{equation}\label{erot}
 \dot{E}_{\rm rot}=-4\pi^2I \frac{\dot{P}}{P^3},
\end{equation}
associated to its spin-down rate $\dot{P}$, where $P$ is the rotational period, and $I$ is
the momentum of inertia.
Assuming that all of the rotational energy lost by the the star is carried away by magnetic dipole
radiation, and equaling Eqs.~(\ref{edot}) and (\ref{erot}) we deduce the expression of pulsar
magnetic dipole moment,
\begin{equation}\label{moment}
 m=\left(\frac{3c^3I}{8\pi^2}P\dot{P}\right)^{1/2}.
\end{equation}

Thus, from Eq.~(\ref{magnetic_moment}) we obtain the surface magnetic field at the equator $B_e$ as (\cite{Ferrari&Ruffini_1969})
\begin{equation}\label{MagneticField}
 B_e=B_p/2=\left(\frac{3c^3I}{8\pi^2R^6}P\dot{P}\right)^{1/2},
\end{equation}
where $P$ and $\dot{P}$ are observed properties and the moment
of inertia $I$ and the radius $R$ of the star model dependent quantities. 
Within the model commonly addressed as the
magnetar one (\cite{Duncan}; \cite{Thompson}), based on a canonical neutron star of $M=1.4M_{\odot}$ and $R=10$
km and then $I\sim 10^{45}\rm{g}$ $\rm{cm^2}$ as the source of SGRs/AXPs, and from Eqs.~(\ref{moment}) and
(\ref{MagneticField}), we obtain the
magnetic dipole moment and the magnetic field
of the neutron star pulsar,
\begin{equation}\label{m_NS}
 m_{\rm NS}=3.2\times10^{37}(P\dot{P})^{1/2} \rm{emu},
\end{equation}
and
\begin{equation}\label{B_NS}
 B_{\rm NS}=3.2\times10^{19}(P\dot{P})^{1/2} \rm{G}.
\end{equation}

We will describe all SGRs and AXPs as white dwarfs of radius $R= 3000$ km and a mass of $M=1.4M_{\odot}$, consistent with recent studies
of fast and
very massive white dwarfs (see \cite{JRueda}). These values
of mass and radius generate a momentum of inertia of $I\sim 1.26\times 10^{50}\rm{g}$ $\rm{cm^2}$,
which we will adopt hereafter in this work as the fiducial white dwarf model parameter.
Using this parameter we obtain the
magnetic dipole moment and the magnetic field
of the white dwarf pulsar,
\begin{equation}\label{m_WD}
 m_{\rm WD}=1.14\times10^{40}(P\dot{P})^{1/2} \rm{emu},
\end{equation}
and
\begin{equation}\label{B_WD}
 B_{\rm WD}=4.21\times10^{14}(P\dot{P})^{1/2} \rm{G}.
\end{equation}

These results clearly shows that the scale of the dipole magnetic moment in a WD is $\sim10^3$ times larger than for neutron stars,
essentially the factor seen in the X-ray luminosity of SGRs/AXPs described as white dwarfs when compared with the $L_X$ of slow neutron star
pulsars as XDINS (see e.g.,~\cite{Kaplan2005};~\cite{Kaplan2009};~\cite{Haberl};~\cite{Malofeev2007}) and high-$B$ pulsars
(see e.g.,~\cite{Espinoza};~\cite{Livingstone2};~\cite{Gavriil};~\cite{Kumar}) 
that have essentially the same period ($P\sim$1 to 10 s) as those of SGRs/AXPs.
Furthermore, the surface magnetic field for WDs is $\sim10^5$ smaller than the ones of neutron stars, eliminating all
the overcritical $B$ fields deduced in the magnetar model.

\section{Observations of SGRs with low-$B$ and white dwarf pulsars}

In the last years, new astronomical observations of two SGRs with low magnetic field challenged the
magnetar model of SGRs where the large magnetic field is
the source of the steady X-ray luminosity observed, and also responsible for the outbursts seen as
the main characteristic of this type of pulsar. Moreover, the characteristic ages $\tau=P/2\dot{P}$
of these two magnetars SGR 0418+5729 and Swift J1822.3-1606 are $\tau \sim (10^6-10^7)$ years, making them older
compared to all other SGRs and AXPs $\tau \sim (10^3-10^4)$ years, an age usually seen as an indication for
the association of SGRs/AXPs with young neutron star produced in 
supernova explosions (see e.g.,~\cite{NandaRea};~\cite{NandaRea2};~\cite{Livingstone}).

These two sources, although old possessing a low surface dipole magnetic field, are still very active,
with an X-ray quiescent luminosity $L_X \sim 200 \dot{E}_{\rm rot}$ (where $\dot{E}$ is calculated with
a neutron star momentum of inertia). To explain this energetic balance in the magnetar model, the toroidal magnetic 
field needs to be two order of magnitude larger than the surface poloidal field (see~\cite{Ciolfi}). 

Furthermore, some sources have been proposed as candidates for white dwarf pulsars. A specific example is AE Aquarii,
the first white dwarf pulsar, very fast with a short period $P= 33.08$ s (\cite{Terada2};~\cite{Angel};~\cite{Ferrario2}). 
The rapid breaking of the white dwarf and the nature of the hard X-ray emission pulse
detected with the Japanese SUZAKU space telescope can be explained in
terms of a spin-powered pulsar mechanism (\cite{Ikhasanov2};~\cite{Ikhasanov3};~\cite{Ikhasanov1}).
Although AE Aquarii is a binary
system with orbital period $\sim9.88$ hr, the power due to the accretion of matter is very likely
inhibited by the fast rotation of the white dwarf (see~\cite{Ikhasanov4}).

More recently, the X-ray multimirror mission (XMM)-Newton satellite had observed a white dwarf 
faster than AE Aquarii. Mereghetti et al. (2009)
showed that the X-ray pulsator RX J0648.0-4418 is a massive white dwarf with 
mass $M=1.28M_\odot$ and radius $R= 3000$ km, with very fast spin period of $P = 13.2$ s, that
belongs to the binary systems HD 49798/RX J0648.0-4418 (see~\cite{Mereghetti3};~\cite{Mereghetti4}).
The luminosity $L_X \sim 10^{32}$ erg/s in this source is produced by accretion
onto the white dwarf of helium-rich matter from the wind of the companion, as discussed
by Mereghetti et al. (2009). In this work, we
do not consider the accretion model. Instead, we
describe RX J0648.0-4418 as
a rotationally powered white dwarf, and obtain the magnetic dipole moment and magnetic field 
using the above mentioned fiducial parameters of a fast rotating magnetized
white dwarf.

EUVE J0317-855, is another observed WD pulsar candidate discovered as an extreme-ultraviolet (EUV)
source by the {\it{ROSAT Wide Field Camera}} and {\it{Extreme Ultraviolet Explorer EUVE}} survey (\cite{Ferrario};\cite{Kulebi}).
Barstow et al. (1995) obtained a period
of $P\sim$ 725 s, which is also a fast, massive and very magnetic white dwarf with a dipole magnetic field of
$B\sim 4.5\times 10^8$ G, and a mass $(1.31 - 1.37)M_\odot$ which is relatively large
compared with the typical WD mass $\sim 0.6 M_\odot$. However, no pulse emission have been detected from
EUVE J0317-855, which may suggest that electron-positron creation
and acceleration does not occur (see~\cite{Kashiyama}). EUVE J0317-855 has a white dwarf companion, but there is supposed to be
no interaction between them because of their large separation ($\gtrsim 10^3$ UA).
\begin{longtable}{llllll}
  \caption{Similarities of SGRs with low magnetic field and white dwarf pulsars. The observational properties of five sources:
SGR 0418+5729, Swift J1822.3-1606, two observed white dwarf pulsars, and the candidate EUVE J0317-855. For
the SGR 0418+5729 and Swift J1822.3-1606 the parameters
$P$, $\dot{P}$ and $L_X$ have been taken from the McGill
online catalog at www.physics.mcgill.ca/~pulsar/magnetar/main.html. The characteristic age is given
by Age $=P/2\dot{P}$ and the WD magnetic moment and
surface magnetic field are calculated by Eqs.~(\ref{m_WD}) and (\ref{B_WD}), respectively.}\label{ta1}
\hline              
& SGR 0418+5729 & Swift J1822.3-1606 & AE Aquarii & RXJ 0648.0-4418 &EUVE J0317-855 \\ 
\hline
\endhead
\hline
\endfoot
\hline
\endlastfoot
\hline
$P$(s) &9.08 &8.44 & 33.08 &13.2 &725\\
$\dot{P}$ $(10^{-14})$ &$<$ 0.6 &8.3 &5.64 &$<$ 90 &- \\
$\rm Age$ $(\rm Myr)$ &24 &1.6 &9.3 &0.23 &- \\
$L_X$ $(\rm erg/s)$ &$\sim6.2\times10^{31}$ &$\sim4.2\times10^{32}$ &$\sim10^{31}$ &$\sim10^{32}$ &- \\
$B_{\rm WD}$(G) &$<$ $9.83\times10^7$ &$3.52\times10^{8}$ &$\sim 5\times10^7$ &$<$ $1.45\times10^9$ &$\sim 4.5\times10^8$ \\
$m_{\rm WD}$(emu) &$2.65\times10^{33}$ &$0.95\times10^{34}$ &$\sim1.35\times10^{33}$ &$3.48\times10^{34}$ &$1.22\times10^{34}$ \\
\end{longtable}
In Table~\ref{ta1} we extend the comparison made in Malheiro, Rueda and Ruffini (2012) of the observational properties
of one low-B SGR with one fast white dwarf,
including the recent observations of one more low-$B$ SGR and two other fast and magnetic white dwarfs.
We see that several features of the two SGRs with low magnetic field are
very similar to those of fast, magnetic white dwarfs recently 
observed. These two SGRs with low-$B$ have a characteristic ages of Myr, comparable with the WDs ages. Furthermore, for the first time, a low surface
magnetic dipole field of $B\sim10^8-10^9$ G is still producing an outbursts: the steady $L_X$ luminosity of these low-$B$ SGRs is still 100 times
larger than their neutron star spin-down luminosity. This unusual phenomenology is not easy to accommodate in the magnetar model, and needs
quite strong toroidal field in the star interior.
However, in the white dwarf model, the magnetic fields of the SGR 0418+5729 and Swift J1822.3-1606 are better understood, since they are of the same order 
as the high magnetic fields of the fast and magnetic white dwarfs $B_{\rm WD}\sim(10^7-10^8)$ G. The steady $L_X$ luminosity of these two SGRs is 
also of the same order, $L_X\sim10^{31}-10^{32}$ erg/s, as the X-ray luminosity obtained for the three magnetic and fast WDs recently observed.
Furthermore, as white dwarfs, their $L_X$ luminosity is well explained by the spin-down luminosity, which is large enough.
We also show that the magnetic dipole moment of these two low-$B$ SGRs is in the same range as those of the fast and magnetic white dwarfs,
$m_{\rm WD}\sim (10^{33}-10^{34})$ emu. In the next section, we will explore more the comparison between the magnetic dipole moment of SGRs/AXPs
and those of polar and isolated white dwarfs.
It is true that RXJ 0648.0-4418 and EUVE J0317-855 have never displayed 
SGR-like X-ray bursting activity. However the case of the white dwarf  
AE Aquarii, it has shown optical and gamma ray burst 
activity (see e.g.,~\cite{Jager};~\cite{Chadwick};~\cite{Jager2};~\cite{Oruru})
and also, hard X-ray pulsations were observed in AE Aquarii by 
the Suzaku satellite (see \cite{Terada2}).
All these strong similarities between the two SGRs with low-$B$ and the magnetic white dwarfs expressed on Table~1, seem to indicate that
some of the SGRs/AXPs, and at least these SGRs with low magnetic field, could be {\it magnetic white dwarfs} and not neutron stars.

\section{Dipole magnetic moment of neutron stars and white dwarfs}

The possibility that strongly magnetized white dwarfs, could behave as neutron star pulsars
with a pulse emission of
high-energy photons in the X-ray to $\gamma$-ray band, gave origin to interesting and very suggestive
plots made for the first time by Terada et al. (2008) (see \cite{Terada3};~\cite{Terada4};~\cite{Terada5}): plots of the magnetic field strength and
the dipole magnetic field as a function of the period of neutron stars and {\it magnetic} white dwarfs.
In this paper, we reproduce these plots including the SGRs and AXPs as neutron stars (magnetars)
and white dwarfs, using for them the fiducial parameters presented before, and including the fast WDs
recently discovered.

\begin{figure}
  \begin{center}
    \FigureFile(80mm,80mm){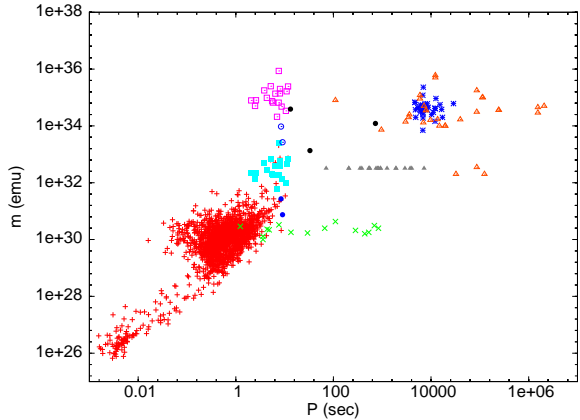}
  \end{center}
  \caption{(Color online) The figure shows the magnetic dipole moment
of neutron stars and magnetic white dwarfs as a function
of the rotational period. The red plus signs designate the
known pulsars from the ATNF online pulsar database. The green cross points represent X-ray pulsars.
The blue full square and the pink square
correspond to SGRs and AXPs as neutron stars and white dwarfs, respectively. The two blue full and open circle points
are the recent observed SGR 0418+5729 and Swift J1822.3-1606 considering neutron stars and white dwarfs parameters, respectively. 
The black full point are the three observed white dwarf pulsar candidates: the X-ray pulsator RX J0648.0-4418,
AE Aquarii and EUVE J0317-855. The gray triangles are the intermediate
polar and the other points are polar and isolated white dwarfs.}\label{figure1}
\end{figure}

In Fig.~\ref{figure1}, the magnetic dipole moment $m$ is presented as a function of
the rotational period. To calculate from Eq.~(\ref{m_WD}) the values of $m$ for SGRs/AXPs (as white dwarfs),
and also for the three fast white dwarfs of Table~\ref{ta1}, we used the radius $R=3000$ km in agreement
with Boshkayev et al. (2012), whereas for the other white dwarfs we adopted
$R=10^4$ km in Eq.~\ref{magnetic_moment}, following Terada \& Dotani (2010) (for NSs the radius used is $R=10$ km). 
The magnetic dipole moment $m$ of the two recently discovered
SGRs (blue full circle) are comparable with those observed for the fast white dwarfs,
also plotted (black full circles). The fast magnetic white dwarfs are separated in two classes: isolated
(orange triangles) and polar (blue asterisks), very magnetic with $B\sim (10^7-10^8)$ G, and
the intermediate polar ones (gray triangle) with weaker field $B\sim 10^5$ G (few isolated magnetic
WDs also belong to that class).
The magnetic WDs shown in Fig.~\ref{figure1},
form the complete sample
of this type of white dwarfs for which the period $P$ and magnetic
field $B$ are known, obtained by the Sloan Digital Sky Survey (SDSS) project.
The catalog from the SDSS, contains about 9 000 objects (\cite{Eisenstein}).
Among them, 480 white dwarfs have high magnetic field strength in the range of $10^4$ to
$10^9$ G (\cite{Kawka}). We selected 82 WDs from this sample, for which
both period $P$ and magnetic field $B$ are known.

This plot also shows that the property of the magnetic dipole moment range $10^{34}{\rm emu}\leq m \leq10^{36}{\rm emu}$ for
isolated and polar white dwarfs, as being almost independent of the rotational period, is also reproduced by the SGRs and
AXPs in the white dwarf model. In fact, the magnetic dipole moment for polar and isolated magnetic white
dwarfs is always in a specific range despite the fact that their rotational period change from $10^4$ to $10^6$ s, a result
that can be seen as part of the phenomenology of magnetic WDs.
This phenomenology is also seen in almost all the SGRs/AXPs when described as white dwarfs: they have much smaller rotational
periods ($P\sim10$ s) than the isolated magnetic white dwarfs and still have magnetic dipole moment essentially in the same range as those of
magnetic white dwarfs (see Fig.~\ref{figure1}).
Furthermore, on Fig.~\ref{figure1} looking at the SGRs/AXPs as WDs and also the polar white dwarfs, we see a gap in period of 10 s to $10^4$ s
between them.
The absence of magnetic WDs in this gap in period supports the formation 
mechanism for the 
SGRs/AXPs already discussed before: they could be formed in binary systems, where one
of the stars is a very magnetic WD with a long period ($10^4$ s or more) that is spun up after the disruption of the system as
discussed in Malheiro, Rueda and Ruffini (2012) or by WDs mergers (see e.g.,~\cite{Rueda2013};~\cite{GBerro2};~\cite{Ilkov2}).
Thus, if this spin-up effect of polar WDs due to the break-up of binary systems (or mergers) is strong enough, it can explain how the usual
long period observed 
for a magnetic white dwarf can bypass the star period range of $10^4-10$ s, and become a much smaller one, as are the periods of SGRs/AXPs. 
This formation mechanism also indicates
a possible origin for the so
small number of SGRs/AXPs, as we already discussed in the introduction, since very magnetic WDs are not very common among the WDs, and binary systems
with one or even two very magnetic white dwarfs are expected to be even rarer.

The near independence of the magnetic dipole moment for magnetic white dwarfs is not seen for neutron star pulsars where $m$ 
is not constant with the period $P$ but increases with it. 
Moreover, we do not see the same slope of $m$ as a function of the star period (see Fig.~\ref{figure1}), when we extrapolate the magnetic dipole
model of normal X-ray pulsars to larger periods in the range $(2<P<12)$ s as those of the SGRs/AXPs described as neutron stars 
(magnetar model): the slope changes abruptly producing
large values of $m$ ($10^{32}\leq m\leq 10^{33}$), at least 3 order of magnitude larger than the normal neutron star pulsars.
This change of the slope can only be explained by the overcritical magnetic fields deduced in the magnetar model, using the same dipole model
applied for normal pulsars: magnetic fields that were never detected through the measure of cyclotron resonance spectral lines,
and which are also larger than the ones that are consistent with their slope extrapolation for higher periods in Fig.~\ref{figure1}.

The fact that magnetic WDs have the same magnetic moment, almost
independent of their rotational period, and NSs
do not, comes essentially from astronomical observations. However, we
can try to find a physical reason for this by looking at
the different origin of magnetic fields of these two classes
of compact stars. The origin of the magnetic fields of white dwarfs
is understood as coming from a fossil field of the progenitors 
that is amplified by the magnetic flux conservation when the star 
progenitor shrinks due to the collapse, or by systems that merge in 
the common envelope phase of pre-Cataclysmic Variables forming a 
population of isolated magnetic white dwarfs with fields 
of $(10^6-10^9)$ G (\cite{Tout}; \cite{Nordhaus}; \cite{Wickramasinghe}). 
In neutron stars, the strong magnetic 
fields are believed to be formed via a dynamo action in 
rapidly rotating proto-neutron stars (\cite{Thompson}). 
Thus, in neutron stars the strength of the magnetic field 
depends intrinsically on their initial periods and on slow 
down effects on their actual periods.

The values for $m\sim (10^{33}-10^{34})$ emu of the two SGRs with low-$B$, are exactly of the same order as those
of the three white dwarf pulsars observed (see Table~\ref{ta1}), and no more than one order as those of magnitude smaller than
the lower values of the observed range for the magnetic dipole moment of magnetic white dwarfs. This result is consistent,
since these sources have smaller magnetic fields when compared to the other SGRs/AXPs.
The large steady X-ray emission $L_X\sim 10^{35}$ erg/s observed in the SGRs/AXPs is now
understood as a consequence of the fast white dwarf rotation ($P\sim 10$ s), since
the magnetic dipole moment $m$ is of the same scale as the one observed for the very magnetic 
and not so fast white dwarfs. 

The results coming from the WDs magnetic dipole moment discussed in this paper are one more evidence suggesting
that SGRs and AXPs belong to a class of very fast and magnetic
massive white dwarfs.

\section{Conclusions}

We have studied the possibility of describing several SGRs/AXPs as belonging to a class of massive, fast-rotating, highly magnetized white dwarfs, following the 
alternative description of SGRs/AXPs proposed recently in Malheiro, Rueda and Ruffini (2012), and showed 
that the values for $m$ obtained for several SGRs and AXPs are in agreement with the observed,
range $10^{34}{\rm emu}\leq m \leq10^{36}{\rm emu}$ of those of isolated and polar magnetic white dwarfs. This result, together with the fact
that for {\it magnetic} white dwarfs $B\sim(10^6-10^8)$ G their magnetic dipole moments are almost independent of the star rotation 
period ($10^{4}\lesssim P \lesssim10^{6} {\rm s}$), a phenomenology not seen in neutron star rotation-powered
pulsars, suggests a possible {\it magnetic} white dwarf nature for several SGRs/AXPs that have much smaller periods ($P\sim 10$ s).
Thus, the astronomical observation that the WD magnetic dipole moment for 
polar and isolated white dwarfs is always in the range indicated before, despite the fact that their rotational period changes from $10^4$ to $10^6$ s,
is an important evidence that suggests the WD nature of SGRs/AXPs, since they share these some properties, a result that constitutes one of
novelties of our work.
Furthermore, we showed that the scale of the dipole magnetic moment in WD is $\sim10^3$ times larger than that for neutron stars,
exactly the factor seen in the X-ray luminosity of SGRs/AXPs described as white dwarfs when compared with the $L_X$ of slow neutron star
pulsars, such as XDINS, and high-$B$ pulsars that have essentially the same period ($P\sim$1 to 10 s) as those of SGRs/AXPs.

We also presented the recent discoveries of SGR 0418+5729 and Swift J1822.3-1606 with low magnetic fields, sharing some properties with the
recently detected fast WD pulsar AE Aquarii and RXJ 0648.0-4418, and the candidate EUVE J0317-855,
to support the WD description of SGRs and AXPs. In Table~1 we compared the parameters of these two SGRs with low magnetic fields described 
in the white dwarf model with three fast white dwarfs. We concluded from Table~1 that several features of the two SGRs with low magnetic fields are
very similar to those of fast, magnetic white dwarfs, for sure one of the most novel aspects of our work.
They are old, with characteristic ages of Myr, low quiescent X-ray
luminosity $L_X \sim (10^{31}-10^{32})$, magnetic fields of $B_{\rm WD}\sim (10^7-10^8)$ G and magnetic dipole moments
of $m_{\rm WD}\sim (10^{33}-10^{34})$ emu. Moreover, the large steady X-ray emission $L_X\sim 10^{35}$ erg/s observed in the SGRs/AXPs is now
well understood as a consequence of the fast white dwarf rotation ($P\sim 10$ s), since the magnetic dipole moment $m$ is of the same scale 
as that observed for the very magnetic and not so fast white dwarfs.

All these findings support the description of some SGRs/AXPs as belonging to a class of very fast and magnetic
massive white dwarfs, in line with important and recent astronomical observations of massive
fast rotating highly magnetized white dwarfs. We encourage future observational campaigns to verify the radius of these sources, in order to
clarify the nature of SGRs/AXPs.

\bigskip

The authors acknowledges the financial support of the Brazilian agency CAPES, CNPq
and FAPESP (S\~{a}o Paulo state agency, thematic project 2007/03633-3).
We are grateful to Y. Terada by the data points of the figure in this paper, and also by the valuable discussions in the 13th Marcel Grossmann Meeting.


\begin{thebibliography}{}
\bibitem[Alcock et al.(1986)]{Olinto} Alcock, C., Farhi, E., \& Olinto, A. 1986, \apj, 310, 261
\bibitem[Allen \& Horvath(2004)]{Allen} Allen, M. P., \& Horvath, J. E. 2004, \apj, 616, 346
\bibitem[Alpar et al.(2011)]{Alpar2011} Alpar, M. A., Ertan, $\rm\ddot{U}$, \& $\rm\c{C}$ali$\rm s$kan, $\rm\c{S}$. 2011, \apjl, 732, L4
\bibitem[Angel et al.(1981)]{Angel} Angel, J. R. P., Borra, E. F., \& Landstreet, J. D. 1981, \apjs, 45, 457
\bibitem[Barstow et al.(1995)]{Barstow} Barstow, M. A., Jordan, S., O'Donoghue, D., Burleigh, M. R., Napiwotzki, R., \& Harrop-Allin, M. K. 1995, \mnras, 277, 971
\bibitem[Beloborodov \& Thompson(2007)]{Beloborodov} Beloborodov, A. M., \& Thompson, C. 2007, \apj, 657, 967
\bibitem[F. Bernardine et al.(2011)]{Bernardine} Bernardine, F. et al. 2011, \aap, 529, A19
\bibitem[Boshkayev et al.(2013)]{JRueda2013} Boshkayev, K., Izzo, L., Rueda, J. A., Ruffini, R. 2013, \aap, 555, A151
\bibitem[Boshkayev et al.(2012)]{JRueda} Boshkayev, K., Rueda, J. A., Ruffini, R., \& Siutsou, I. 2012, \apj, 762, 117
\bibitem[Ciolfi et al.(2009)]{Ciolfi} Ciolfi, R., Ferrari, V., Gualtieri, L. \& Pons, J. A. 2009, \mnras, 397, 913
\bibitem[Chadwick et al.(1995)]{Chadwick} Chadwick, P. M. et al. 1995, Astroparticle Phys., 4, 99
\bibitem[Coelho \& Malheiro(2012)]{Coelho} Coelho, J. G., \& Malheiro, M. 2012, Int. J. Mod. Phys. Conf. Ser., 18, 96
\bibitem[Coelho \& Malheiro(2013)]{Coelho2} Coelho, J. G., \& Malheiro, M. 2013, AIP Conference Proceedings, 1520, 258
\bibitem[Coelho \& Malheiro(2013)]{jaziel} Coelho, J. G., \& Malheiro, M. 2013, arXiv:1303.0863v2
\bibitem[Jager \& Meintjes(1993)]{Jager} de Jager, O. C. \& Meintjes, P. J. 1993, \aap, 268, L1-L4
\bibitem[Dey et al.(1998)]{Dey1998} Dey, M., Bombaci, I., Dey, J., Ray, S., \& Samanta, B. C. 1998, Phys. Lett. B, 438, 123
\bibitem[Duncan \& Thompson(1992)]{Duncan} Duncan, R. C., \& Thompson, C. 1992, \apj, 392, L9
\bibitem[Eatough et al.(2013)]{Nature2013} Eatough, R. P. et al. 2013, \nat, doi: 10.1038/nature12499
\bibitem[Eisenstein et al.(2006)]{Eisenstein} Eisenstein, D. J. et al. 2006, \apjs, 167, 40
\bibitem[Espinoza et al.(2011)]{Espinoza} Espinoza, C. M., Lyne, A. G., Kramer, Manchester, M. R. N., \& Kaspi, V. M. 2011, \apj, 741, L13
\bibitem[Ferrari \& Ruffini(1969)]{Ferrari&Ruffini_1969} Ferrari, A., \& Ruffini, R. 1969, ApJL, 158, L71+
\bibitem[Ferrario et al.(1997)]{Ferrario} Ferrario, L., Vennes, S., Wickramasinghe, D. T., Bailey, J. A., \& Christian, D. J. 1997, \mnras, 292, 205
\bibitem[Ferrario \& Wickramasinghe(2005)]{Ferrario2} Ferrario, L., \& Wickramasinghe, D. T. 2005, \mnras, 356, 615
\bibitem[Ferrario \& Wickramasinghe(2006)]{Ferrario2006} Ferrario, L., \& Wickramasinghe, D. T. 2006, \mnras, 367, 1323
\bibitem[Gabler et al.(2012)]{Gabler2012} Gabler, M., Cerd\'{a}-Dur\'{a}n, P., Stergioulas, N.,  Font, J. A. \& M$\rm\ddot{u}$ller, E. 2012, \mnras, 421, 2054
\bibitem[Gabler et al.(2013)]{Gabler2013} Gabler, M., Cerd\'{a}-Dur\'{a}n, P., Font, J. A., M$\rm\ddot{u}$ller, E. \& Stergioulas, N. 2013, \mnras, 430, 1811
\bibitem[Gaensler et al.(2001)]{Gaensler} Gaensler, B. M., Slane, P. O., Gotthelf, E. V. \& Vasisht, G. 2001, \apj, 559, 963
\bibitem[Garc\'{i}a-Berro et al.(2012)]{GBerro2} Garc\'{i}a-Berro et al. 2012, \apj, 749, 25
\bibitem[Garc\'{i}a-Berro et al.(2012)]{GBerro} Garc\'{i}a-Berro et al. 2012, arXiv:1209.2622v1
\bibitem[Gavriil et al.(2008)]{Gavriil} Gavriil, F. P., Gonzalez, M. E., Gotthelf, E. V., Kaspi, V. M., Livingstone, M. A., Woods, P. M. 2008, Science, 319, 1802
\bibitem[Gelfand \& Gaensler(2007)]{Gelfand} Gelfand, J. D., \& Gaensler, B. M. 2007, \apj, 667, 1111
\bibitem[Haberl(2007)]{Haberl} Haberl, F. 2007, \apss, 308, 181
\bibitem[Horvath(2007)]{Horvath2007} Horvath, J. E. 2007, Astrophys. Space Sci., 308, 431
\bibitem[Ikhsanov(1998)]{Ikhasanov2} Ikhsanov, N. R. 1998, \aap, 338, 521
\bibitem[Ikhsanov \& Biermann(2006)]{Ikhasanov3} Ikhsanov, N. R., \& Biermann, P. L. 2006, \aap, 445, 305
\bibitem[Ikhsanov \& Beskrovnaya(2008)]{Ikhasanov1} Ikhsanov, N. R., \& Beskrovnaya, N. G. 2008, astro-ph: 0809.1169v1
\bibitem[Ikhsanov \& Beskrovnaya(2012)]{Ikhasanov4} Ikhsanov, N. R., \& Beskrovnaya, N. G. 2012, Astronomy Reports, 56, 595
\bibitem[Ilkov \& Soker(2012)]{Ilkov} Ilkov, M., \& Soker, N. 2012,\mnras, 419, 1695
\bibitem[Ilkov \& Soker(2012)]{Soker2} Ilkov, M., \& Soker, N. 2012, arXiv:1208.0953
\bibitem[Ilkov \& Soker(2013)]{Ilkov2} Ilkov, M., \& Soker, N. 2013, \mnras, 428, 579
\bibitem[Itoh(1970)]{Itoh1970} Itoh, N. 1970, Prog. Theor. Phys., 44, 291
\bibitem[Kaplan \& van Kerkwijk(2005)]{Kaplan2005} Kaplan, D. L., \& van Kerkwijk, M. H. 2005, \apj, 628, L45
\bibitem[Kaplan \& van Kerkwijk(2009)]{Kaplan2009} Kaplan, D. L., \& van Kerkwijk, M. H. 2009, \apj, 692, L62
\bibitem[Kasen \& Bildsten(2010)]{kasen} Kasen, D., \& Bildsten, L. 2010, \apj, 717, 245
\bibitem[Kashiyama et al.(2011)]{Kashiyama} Kashiyama, K., Ioka, K., \& Kawanaka, N. 2011, \prd, 83, 023002
\bibitem[Kaspi et al.(2003)]{Kaspi} Kaspi, V. M., Gavriil, F. P., Woods, P. M., Jensen, J. B., Roberts, M. S. E., \& Chakrabarty, D. 2003, \apj, 588, L93
\bibitem[Kawka et al.(2007)]{Kawka} Kawka, A., Vennes, S., Schmidt, G. D., \& Wickramasinghe, D. T. 2007, \apj, 654, 499
\bibitem[Kennea et al.(2013)]{Kennea} Kennea, J. A. et al., 2013, \apjl, 770, L24
\bibitem[Kepler et al.(2012)]{Kepler} Kepler, S. O. et al., 2013, \mnras, 429, 2934
\bibitem[K$\rm \ddot{u}$lebi et al.(2009)]{Kulebi2009} K$\rm \ddot{u}$lebi, B., Jordan, S., Euchner, F., G$\rm \ddot{a}$nsicke, B. T., \& Hirsch, H. 2009, 
\aap, 506, 1341
\bibitem[K$\rm \ddot{u}$lebi et al.(2010)]{Kulebi} K$\rm \ddot{u}$lebi, B., Jordan, S., Nelan, E., Bastian, U., \& Altmann, M. 2010b, \aap, 524, A36
\bibitem[Kumar \& Safi-Harb(2008)]{Kumar} Kumar, H. S., \& Safi-Harb, S. 2008, \apj, 678, L43
\bibitem[Liebert et al.(1983)]{Liebert1983} Liebert, J., Schmidt, G. D., Green, R. F., Stockman, H. S., \& McGraw, J. T. 1983,
\apj, 264, 262
\bibitem[Livingstone et al.(2011)]{Livingstone2}  Livingstone, M. A.,Ng, C.-Y., Kaspi, V. M., Gavriil, F. P., \& Gotthelf, E. V. 2011, \apj, 730, 66
\bibitem[Livingstone et al.(2011)]{Livingstone} Livingstone, M. A., Scholz, P., Kaspi, V. M., Ng., C.-Y., \& Gavriil, F. P. 2011, \apj, 743, L38
\bibitem[Malheiro et al.(2003)]{MMalheiro2003} Malheiro, M., Fiolhais, M. \& Taurines, A. R. 2003, Journal of Physics. G, 29, 1045
\bibitem[Malheiro et al.(2012)]{MMalheiro} Malheiro, M., Rueda, J. A., \& Ruffini, R. 2012, \pasj, 64, 56
\bibitem[Malofeev et al.(2007)]{Malofeev2007} Malofeev, V. M., Malov, O. I., Teplykh, D. A. 2007, \apss, 308, 211
\bibitem[Manchester et al.(2005)]{Manchester} Manchester, R. N., Hobbs, G. B., Teoh, A., \& Hobbs, M. 2005, \aj, 129, 1993
\bibitem[Jager \& Meintjes(2000)]{Jager2} Meintjes, P. J., \& de Jager, O. C. 2000, \mnras, 311, 611
\bibitem[Mereghetti(2008)]{Mereghetti} Mereghetti, S. 2008, \aapr, 15, 225
\bibitem[Mereghetti et al.(2009)]{Mereghetti3} Mereghetti, S., Tiengo, A., Esposito, P., La Palombara, N., Israel, G. L., \& Stella, L. 2009, Science, 325, 1222
\bibitem[Mereghetti et al.(2011)]{Mereghetti4} Mereghetti, S. et al. 2011, \apj, 737, 51
\bibitem[Mori et al.(2013)]{Kaya} Mori, K. et al., 2013, \apjl, 770, L23
\bibitem[Morini et al.(1988)]{Morini} Morini, M., Robba, N. R., Smith, A., \& van der Klis, M. 1988, \apj, 333, 777
\bibitem[Negreiros et al.(2009)]{Negreiros2009} Negreiros, R., Weber, F., Malheiro, M. \&  Usov, V. 2009, \prd, 80, 083006
\bibitem[Nordhaus et al.(2011)]{Nordhaus} Nordhaus, J., Wellons, S., Spiegel, D. S., Metzger, B. D., \& Blackman, E. G. 2011, Proceedings
of the National Academy of Science, 108, 3135
\bibitem[Oruru \& Meintjes(2012)]{Oruru} Oruru, B., \& Meintjes, P. J. 2012, \mnras, 421, 1557
\bibitem[Paczynski(1990)]{Paczynski} Paczynski, B. 1990, \apjl, 365, L9
\bibitem[Paulucci \& Horvath(2008)]{Paulucci} Paulucci, L. \& Horvath, J. E. 2008, \prc, 78, 064907 
\bibitem[Rea et al.(2008)]{NandaRea2008} Rea, N., Zane, S., Turolla, R., Lyutikov, M., \& G$\ddot{o}$tz, D. 2008, \apj, 686, 1245
\bibitem[Rea et al.(2010)]{NandaRea} Rea, N. et al., 2010, Science, 330, 944
\bibitem[Rea et al.(2012)]{Turolla} Rea, N., Pons, J. A., Torres, D. F., \& Turolla, R. 2012, \apj, 748, L12
\bibitem[Rea et al.(2012)]{NandaRea2} Rea, N. et al., 2012, \apj, 754, 27
\bibitem[Rueda et al.(2013)]{Rueda2013} Rueda, J. A., et al. 2013, \apj, 772, L24
\bibitem[Schmidt et al.(1986)]{Schmidt1986} Schmidt, G. D., West, S. C., Liebert, J., Green, R. F., \& Stockman, H. S. 1986, \apj, 309, 218
\bibitem[Schmidt et al.(1992)]{Schmidt1992} Schmidt, G. D., Bergeron, P., Liebert, J., \& Saffer, R. A. 1992, \apj, 394, 603
\bibitem[Shapiro \& Teukolsky(1983)]{Shapiro} Shapiro, S. L., \& Teukolsky, S. A. 1983, Black holes, white dwarfs, and neutron stars:
The physics of compact objects, New York: Wiley-Interscience
\bibitem[Silvestri et al.(2007)]{Silvestri} Silvestri, N. M. et al. 2007, \apj, 134, 741
\bibitem[Terada et al.(2008c)]{Terada2} Terada, Y. et al., 2008c, \pasj, 60, 387
\bibitem[Terada et al.(2008d)]{Terada3} Terada, Y. et al., 2008d, Advances in Space Research, 41, 512
\bibitem[Terada et al.(2008)]{Terada4} Terada, Y. et al., 2008, ESA Special Publication, 622, 521
\bibitem[Terada \& Dotani(2010)]{Terada5} Terada, Y., \& Dotani, T. 2010, astro-ph.HE: 1006.5274v1
\bibitem[Terada(2013)]{Terada2013} Terada, Y. 2013, arXiv:1306.4053
\bibitem[Thompson \& Duncan(1995)]{Thompson} Thompson, C., \& Duncan, R. C. 1995, \mnras, 275, 255
\bibitem[Tong \& Song(2010)]{Tong2010} Tong, H., Song, L. M., \& Xu, R. X. 2010, \apj, 725, L196
\bibitem[Tong \& Xu(2011)]{Tong2011} Tong, H., \& Xu, R. X. 2011, Int. J. Mod. Phys. E, 20, 15
\bibitem[Tout et al.(2008)]{Tout} Tout, C. A., Wickramasinghe, D. T., Liebert, J., Ferrario, L., \& Pringle, J. E. 2008, \mnras, 387, 897.
\bibitem[Tr$\rm\ddot{u}$mper et al.(2013)]{Trumper} Tr$\rm\ddot{u}$mper, J. E., Denner, K., Kylafis, N. D.,
Ertan, $\rm\ddot{U}$ \& Zezas, A. 2013, \apj, 764, 49
\bibitem[G. Vasisht \& E. V. Gotthelf(1997)]{Vasisht} Vasisht, G., \& Gotthelf, E. V. 1997, \apj, 486, L129
\bibitem[Vink \& Kuiper(2006)]{Vink2006} Vink, J., \& Kuiper, L. 2006, \mnras, 370, L14
\bibitem[Vink(2008)]{Vink2008} Vink, J. 2008, Adv. Space Res., 41, 503
\bibitem[Wickramasinghe \& Ferrario(2000)]{Wickramasinghe} Wickramasinghe, D. T., \& Ferrario, L. 2000, \pasp, 112, 873
\bibitem[Woods et al.(2004)]{Woods} Woods, P. M. et al., 2004, \apj, 605, 378
\bibitem[Xu(2007)]{Xu2007}Xu, R. 2007, Advances in Space Research, 40, 1453
\bibitem[Zhang \& Gil(2005)]{Zhang} Zhang, B., \& Gil, J. 2005, \apj, 631, L143

\end{thebibliography}
\end{document}